# Bilayer Vanadium Dioxide Thin Film with Elevated Transition Temperatures and High Resistance Switching

Achintya Dutta,[1, a)] Ashok P.,[2,*] and Amit Verma[2]

[1] *Department of Condensed Matter Physics and Materials Science, Tata Institute of Fundamental Research, Mumbai 400005, India.*

[2] *Department of Electrical Engineering, Indian Institute of Technology Kanpur, Kanpur 208016, India.*

* Corresponding Author, correspondence to: Semiconductor Devices Lab, Department of Electrical Engineering, Indian Institute of Technology, Kanpur 208016, India.

Email Addresses: a.dutta.nano@gmail.com (Achintya Dutta), ashoksnp90@gmail.com (Ashok P.), amitkver@iitk.ac.in (Amit Verma)

Despite widespread interest in the phase-change applications of vanadium dioxide ($VO_2$), the fabrication of high-quality $VO_2$ thin films with elevated transition temperatures ($T_{IMT}$) and high Insulator-Metal-Transition resistance switching still remains a challenge. This study introduces a two-step atmospheric oxidation approach to fabricate bilayer $VO_{2-x}/VO_2$ films on a c-plane sapphire substrate. To quantify the impact of the $VO_2$ buffer layer, a single-layer $VO_2$ film of the same thickness was also fabricated. The bilayer $VO_{2-x}/VO_2$ films wherein the top $VO_{2-x}$ film was under-oxidized demonstrated an elevation in $T_{IMT}$ reaching ~97 °C, one of the highest reported to date for $VO_2$ films and is achieved in a doping-free manner. Our results also reveal a one-order increase in resistance switching, with the optimum bilayer $VO_2/VO_2$ film exhibiting ~3.6 orders of switching from 25 °C to 110 °C, compared to the optimum single-layer $VO_2$ reference film. This is accompanied by a one-order decrease in the on-state resistance in its metallic phase. The elevation in $T_{IMT}$, coupled with increased strain extracted from the XRD characterization of the bilayer film, suggests the possibility of compressive strain along the c-axis. These $VO_{2-x}/VO_2$ films also demonstrate a significant change in the slope of their resistance vs temperature curves contrary to the conventional smooth transition. This feature was ascribed to the rutile/monoclinic quasi-heterostructure formed due to the top $VO_{2-x}$ film having a reduced $T_{IMT}$. Our findings carry significant implications for both the lucid fabrication of $VO_2$ thin film devices as well as the study of phase transitions in correlated oxides.

## I. INTRODUCTION

Vanadium dioxide ($VO_2$) is a Mott-Peierls insulator that undergoes Insulator-Metal-Transition (IMT) accompanied by a Structural Phase Transition (SPT) from low-temperature insulating monoclinic (*M1*, $P2_1/c$) to high-temperature metallic tetragonal rutile (*R*, $P4_2/mnm$) phase around 68 °C (~341 K) under ambient pressure[1]. The monoclinic (*M1*) and tetragonal rutile (*R*) structures are significantly different, with the *M1* phase having double the unit cell volume as compared to the *R* phase[2]. The formation of V−V dimers takes place along the c-axis during the *R* to *M1* phase transition[2]. This structural transition is accompanied by a temperature-driven change of several orders of magnitude in the resistance[2], as well as the optical reflectance[3]. These properties not only make $VO_2$ a promising material for devices such as neuromorphic oscillators[4], steep-slope phase FETs[5], volatile memory[6], and optical metasurfaces[7] but also ensure that $VO_2$ serves as a sandbox for studying the

---

[a)] This research was performed while Achintya Dutta was affiliated with Manipal Institute of Technology, Manipal, Karnataka 576104, India.

evolving physics underlying electronic correlation, transport, and phase-transition time-scales in Mott insulators[8]. Recent efforts have focused on harnessing this emergent correlated physics to realize neurobiological systems[9] and fabricate high-frequency RF circuits[10]. To ensure the widespread adoption of VO$_2$ films in the fabrication of the aforementioned devices, a triad of requirements must be met: firstly, a simple methodology to fabricate pure-phase VO$_2$ films that involve a limited thermal budget while attaining low on-state resistance values in their metallic rutile phase; secondly, increasing the magnitude of resistance switching without incurring significant trade-offs; and finally, the development of doping-free techniques to manipulate IMT transition temperatures. Our work in this paper aims to address these pertinent challenges.

Vanadium (V) exhibits multiple valence states, out of which +5, +4, +3, and +2 are commonly observed. This gives rise to at least 22 stable phases of the V−O system, such as VO$_2$, V$_2$O$_3$, V$_2$O$_5$, V$_6$O$_{13}$, and V$_3$O$_5$[11]. These pose a significant challenge in the growth of pure-phase VO$_2$ thin films and normally require exacting control over environmental pressure and gas concentrations. Contrary to the traditional methods reported in the literature, such as chemical vapor deposition (CVD)[12], atomic layer deposition (ALD)[13], molecular beam epitaxy (MBE)[14-16], and pulsed laser deposition (PLD)[17-18], we previously reported a novel, simple methodology involving the atmospheric pressure thermal oxidation (APTO) of sputtered V films on a c-plane sapphire substrate (Al$_2$O$_3$) to achieve pure-phase VO$_2$ films[19]. This decreased the oxidation duration and thermal budget required and allowed the oxidation to be carried out under atmospheric pressure while attaining around three orders of resistance switching. However, for two main reasons, low on-state resistance values could not be attained in these VO$_2$ films. Firstly, due to the low thermal budget, smaller grain sizes were observed in the films, which may result in undesirable high on-state resistance values[17]. Secondly, the lattice parameter mismatch between the Al$_2$O$_3$ substrate ($a = b = 4.75$ A°, $c = 12.99$ A°) and the VO$_2$ film ($a = b = 4.54$ A°, $c = 2.88$ A°) may degrade the overall film quality[20].

Apart from high IMT resistance switching and low on-state resistance, different phase-change applications demand varied transition temperatures. While recent reports have emphasized the fabrication of VO$_2$ films with low transition temperatures (30 - 40 °C) for thermochromic smart window applications[21], it is imperative to note that several specialized devices require high operating temperatures exceeding the bulk VO$_2$ phase transition temperature of ~68 °C. This includes electronics employed in commercial and defense applications, which need to operate above 85 °C and 125 °C, respectively[22], high-temperature micro-bolometers operating around 95 °C[23], and high-temperature electronics used in data centers[24]. The field of reconfigurable RF devices has also emerged as a crucial area, garnering substantial interest from the scientific community, especially in relation to phase-change materials possessing high transition temperatures[10]. However, RF devices like RF switches and oscillators necessitate a minimum transition temperature of ~80 °C, rendering conventional VO$_2$ films unsuitable[10]. For all the aforementioned applications, increased transition temperatures are normally achieved by the stoichiometric doping of $VO_2$ films with elements such as $Fe^{25}$, $Cr^{26}$, and $Ge^{27}$. Although these reports demonstrate elevated transition temperatures, a clear monotonic decrease in the magnitude of IMT resistance switching is often observed with the increase in doping percentage. The Ge-doped $Ge_xV_{(1-x)}O_2$ film reported by A. Krammer et al.[27] with the highest doping concentration demonstrated only ~1.5 orders of resistance switching. This decrease in switching with increasing doping is also evident in the Cr-doped $Cr_xV_{(1-x)}O_2$ films reported by B. L. Brown et al.[26], which demonstrate a slight decrease in the difference between the carrier concentrations in the insulating and metallic phases, implying a reduction in the IMT resistance switching. Thus, achieving elevated transition temperatures while also maintaining or improving IMT resistance switching is a pertinent challenge in the fabrication of phase-change devices[28].

Several studies have documented the utilization of buffer layers or virtual substrates to fabricate high-quality $VO_2$ films devoid of doping in the quest to achieve desirable properties. For instance, E. Breckenfeld et al.[28] introduced a $TiO_2$ buffer layer for the growth of $VO_2$ films on an m-cut sapphire substrate, resulting in a lowered transition temperature of 44 °C due to partial strain between the $TiO_2$ buffer layer and the $VO_2$ film. Similarly, H. Kim et al.[29] investigated a $VO_2/SnO_2/Al_2O_3$ system that relied on epitaxial strain between the $VO_2$ film and the $SnO_2$ buffer to modulate the IMT transition temperature. C. Kang et al.[32] demonstrated the use of a $TiO_2$ buffer layer on a soda lime glass substrate to enhance photoelectric properties. However, most studies employing buffer layers mainly focus on reducing transition temperatures and are often accompanied by a decrease in the magnitude of IMT resistance switching. Only a handful of studies have explored the use of buffer layers such as (110) $TiO_2$[33] to elevate the transition temperature of $VO_2$ films or enhance the magnitude of IMT resistance switching using $Al_2O_3$[30] or yttria-stabilized zirconia[31] buffer layers. Although scarce, these reports suggest that the high crystallinity and improved lattice parameter matching between the buffer layer and the $VO_2$ film contribute to overall film quality improvement. Furthermore, a few bilayer systems have also been employed to gain fundamental insights into the switching dynamics and sub-nanosecond time scales of the phase transitions occurring in $VO_2$[34].

This study investigates bilayer $VO_{2-x}/VO_2$ films fabricated on a c-plane sapphire substrate, employing a two-step atmospheric pressure thermal oxidation (APTO) technique with $VO_2$ acting as a buffer layer. These $VO_{2-x}/VO_2$ films are oxidized for varying durations and subsequently characterized by Raman Spectroscopy and four-probe measurements to pinpoint the optimum oxidation duration that yields pure-phase bilayer $VO_2/VO_2$ films. A single-layer $VO_2$ film of a comparable thickness was synthesized using conventional single-step APTO as a reference for comparison. The optimum bilayer $VO_2/VO_2$ film showed a nearly one-order increase in IMT resistance switching compared to its single-layer counterpart. In addition, a notable one-order reduction in on-state resistance in the metallic rutile phase was also observed. Aside from the optimum $VO_2/VO_2$ film, the under-oxidized $VO_{2-x}/VO_2$ films demonstrated increased transition temperatures between 81.6 °C and 96.9 °C, which was tuneable by varying the oxidation duration of the top $VO_{2-x}$ film. This is among the highest reported transition temperatures for $VO_2$ films and is also achieved in a completely doping-free manner. Moreover, the under-oxidized $VO_{2-x}/VO_2$ films also demonstrate a significant change in the slope of the resistance vs temperature plots during the phase transition, contrary to the conventional smooth phase transition observed in $VO_2$.

## II. MATERIALS AND METHODS

### A. Thin Film Fabrication

To synthesize the bilayer $VO_{2-x}/VO_2$ films, a two-step APTO of DC sputtered V films was carried out as illustrated in Fig. 1. Initially, a c-plane sapphire substrate was sonicated in acetone and isopropyl alcohol (IPA) for ten minutes each, followed by blow drying using dry nitrogen. A 40 nm metallic V film was deposited on these cleaned substrates using DC sputtering. A V sputtering target with a diameter of four inches and a purity of 99.95% was used for the same. Prior to the deposition, the sputtering chamber was pumped to a base pressure of 0.5 mPa. Following this, Ar gas was released into the chamber to maintain a deposition pressure of 3 Pa. Throughout the deposition, the DC power was maintained at 90 W. Sputtering was carried out for four minutes to achieve a film thickness of 40 nm[19]. These V sputtered c-plane sapphire substrates were diced into six pieces. These were placed on a hot plate maintained at 450 °C and oxidized for different durations ($T_{oxd}$) under atmospheric pressure to obtain six $VO_{2-x}$ films with varying degrees of oxidation. To accurately control oxidation times, each sample was quenched on a metallic cold plate immediately after heating[19]. Four-probe temperature-dependent resistance measurement was utilised to quantify the IMT resistance switching ($R_{25 °C}/R_{110 °C}$) for all six samples. IMT resistance switching is defined as the ratio of the resistance of the sample at 25 °C ($R_{25 °C}$) and at 110 °C ($R_{110 °C}$). The sample with the highest magnitude of switching was identified as the pure-phase $VO_2$ film with an optimum oxidation time ($T^*_{oxd}$) out of all the $VO_{2-x}$ films obtained[19]. This optimum 80 nm $VO_2$ film was utilized as the buffer layer for further deposition. A second round of DC sputtering was carried out to deposit a 40 nm metallic V film on top of this buffer layer. This c-plane sapphire substrate with a $VO_2$ buffer layer and a metallic V top film was further diced into six pieces and again oxidized at different $T_{oxd}$. Thus, six samples of bilayer $VO_{2-x}/VO_2$ films were obtained. Raman spectroscopy and four-probe temperature-dependent resistance measurements were repeated on all the samples to identify the pure-phase bilayer $VO_2/VO_2$ film with the highest resistance switching[19].

To fabricate a control sample for comparison, a single-layer $VO_2$ film of the same 160 nm thickness was synthesized as the reference film using the conventional single-step APTO process. An 80 nm metallic V film was initially deposited on a fresh c-plane sapphire substrate, followed by dicing and APTO at 450 °C for different $T_{oxd}$ to obtain single-layer $VO_{2-x}$ films. Out of these, the optimum single-layer $VO_2$ film was identified and selected as our reference film. Optical images of all the samples are presented in Fig. 2.

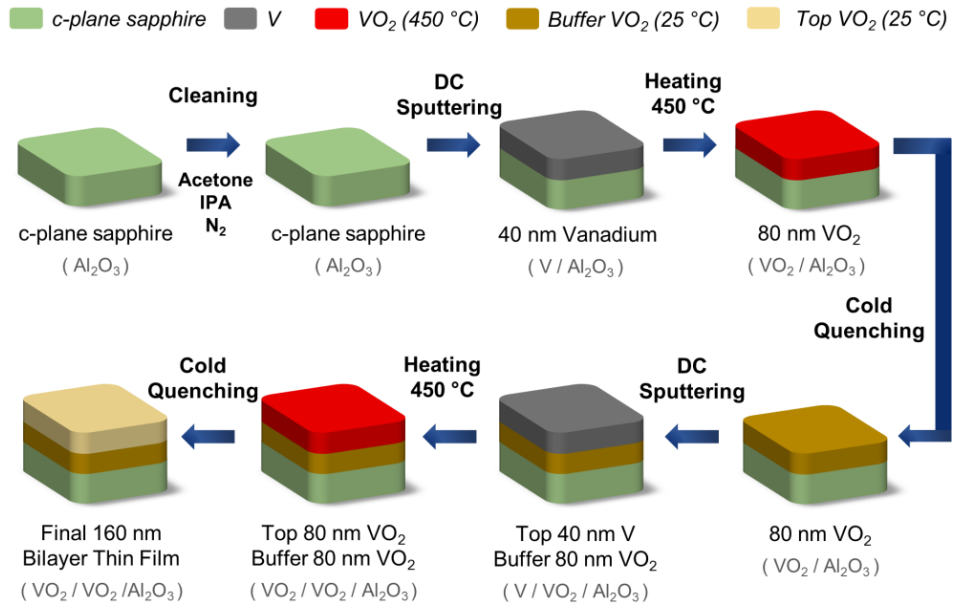

FIG. 1. Two-step atmospheric pressure thermal oxidation (APTO) methodology used to fabricate bilayer $VO_2/VO_2$ films on a c-plane sapphire substrate.

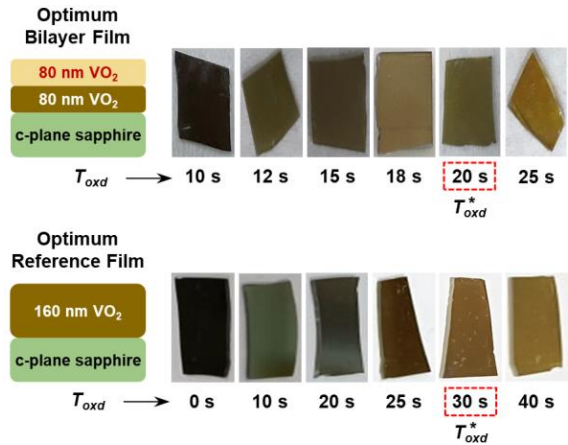

FIG. 2. Graphical illustration of the optimum bilayer and single-layer reference films accompanied by the optical images of all the samples oxidized for different durations.

## B. Characterization Methods

All the temperature-dependent resistance measurements mentioned above were carried out using a four-point probe setup coupled with a temperature-controlled oven in a range of 25 °C to 120 °C. A constant current of 10 µA was applied across the outer probes, while the voltage across the inner probes was measured to calculate the resistance of the films. The crystallinity and phase formation of all the samples were characterized using Raman Spectroscopy with an excitation laser of 532 nm wavelength (Acton Research Corporation Spectra Pro 2500i). X-Ray Diffraction (XRD) using Cu Kα radiation (Rigaku Miniflex Pro) was used to characterize the crystallite planes. To further probe the crystallite size and strain in both the bilayer $VO_2/VO_2$ film, as well as the single-layer $VO_2$ reference film, the Scherrer formula (Eq. 1) and the Williamson-Hall (W-H) formula (Eq. 2) were employed[35]:

$$D_s = \frac{k \cdot \lambda}{B_r \cdot \cos(\theta)} \tag{1}$$

$$B_r \cdot \cos(\theta) = \frac{k \cdot \lambda}{D_{WH}} + \eta \cdot \sin(\theta) \tag{2}$$

Here, $k$ represents the Scherrer constant (0.94), λ represents the wavelength of the X-ray radiation (0.154060 nm), $B_r$ represents the full-width half-maxima (FWHM) of the respective diffraction peaks, $\theta$ represents the Bragg's angle, $D_s$ represents the crystallite size found using the Scherrer formula (Eq. 1), whereas, $D_{WH}$ represents the crystallite size found using the W-H formula (Eq. 2). The strain in the films is represented by $\eta$. In the W-H formula, $D_{WH}$ and $\eta$ are found by plotting $B_r \cdot \cos(\theta)$ vs $\sin(\theta)$, fitting the data linearly, and extracting the intercept and slope of the plot, respectively.

## II. RESULTS AND DISCUSSIONS

Prior to the sputtering of the top V layer to form the bilayer films, the IMT resistance switching ratios of the buffer layer at different $T_{oxd}$ were measured to help identify the optimum $VO_2$ film. As illustrated in Fig. 3 (a), these samples demonstrate a typical behavior with a low magnitude of IMT resistance switching (~one order) initially up to $T_{oxd}$ = 10 s[19]. This is ascribed to the high V content in these samples due to partial oxidation. A narrow window can be observed at $T^*_{oxd}$ = 12 s which shows a maximum resistance switching magnitude of 2.2 orders. A significant decrease in switching can be observed following $T^*_{oxd}$ = 12 s with samples having $T_{oxd}$ = 14 s, 17 s, and 20 s demonstrating decreasing switching magnitudes of 1.8, 1.3, and 0.2 orders respectively due to the gradual increase of $V_2O_5$ content in these films. Thus, we conclude that the sample with $T^*_{oxd}$ = 12 s is the closest to the desired pure-phase $VO_2$ as it demonstrates the highest magnitude of resistance switching. This sample is selected as the buffer layer. As described previously, a second round of 40 nm V was sputtered, followed by dicing, and APTO at different $T_{oxd}$ was repeated to fabricate the top $VO_{2-x}$ film, thus completing the bilayer $VO_{2-x}/VO_2$ films oxidized for different durations.

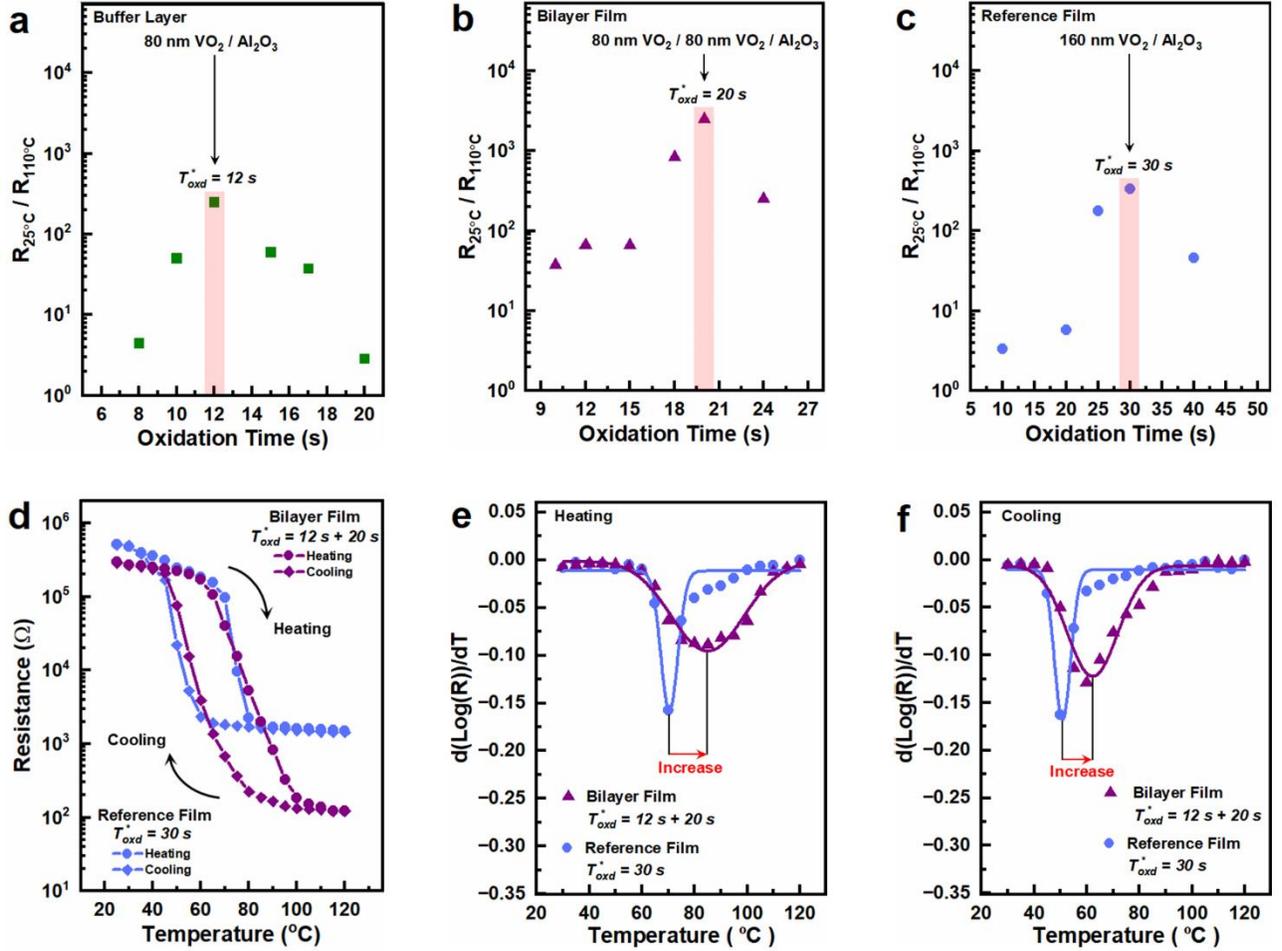

FIG. 3. The insulator-metal phase transition induced resistance switching ratios as a function of oxidation duration for: (a) Buffer film prior to top layer deposition, (b) Bilayer film, (c) Single-layer reference film. (d) IMT resistance switching comparison between the optimum $VO_2/VO_2$ bilayer film and the optimum single-layer film. Derivative of $Log_{10}\{R(T)\}$ for: (e) Heating cycle, (f) Cooling cycle.

Amongst these, to further identify the optimum bilayer $VO_2/VO_2$ film, the IMT resistance switching of all the $VO_{2-x}/VO_2$ films are presented in Fig. 3 (b). The samples oxidized for very short durations, such as $T_{oxd}$ = 9 s, 12 s, and 15 s, show moderately good switching of ~1.3, ~1.9, and ~2 orders, respectively. During APTO of the top V film, the presence of the $VO_2$ buffer layer may cause oxygen diffusion from both the atmosphere, as well as the buffer layer, resulting in faster oxidation of the top film and reasonably low V content even with short durations of oxidation. This is in agreement with the Raman characterization presented in the following paragraphs. The bilayer film with $T^*_{oxd}$ = 20 s demonstrates ~3.6 orders of IMT resistance switching and is clearly the highest amongst all the bilayer films. Thus, we conclude that the sample with $T^*_{oxd}$ = 20 s is the closest to a pure-phase bilayer $VO_2/VO_2$ film and is further confirmed by Raman spectroscopy in the following sections. Moreover, a wider oxidation window is also observed, with samples having $T_{oxd}$ = 18 s and 24 s also showing reasonably high values of ~3 and ~2.5 orders, respectively. To carry out a quantitative comparison, the IMT resistance switching measurements for the single-layer $VO_2$ reference film with a comparable thickness are presented in Fig. 3 (c). We observe a typical behavior with samples oxidized for short durations showing less than one order of switching with a gradual increase peaking at $T^*_{oxd}$ = 30 s which demonstrates ~2.7 orders of IMT resistance switching and can be identified as the optimum single-layer VO$_2$ reference film. A stark improvement of close to one order is visible, with the optimum bilayer film demonstrating ~3.6 orders of IMT resistance switching compared to the optimum single-layer reference film demonstrating ~2.7 orders of switching. To further enunciate this one-order improvement, the resistance as a function of temperature between 25 °C and 120 °C for both films is illustrated in Fig. 3 (d). These results clearly indicate a significant improvement in IMT resistance switching with the insertion of a VO$_2$ buffer layer via two-step APTO. To extract IMT parameters such as transition temperature ($T_{IMT}$), transition interval ($\Delta T_{Tr}$), and transition hysteresis ($\Delta T_{Hys}$), the first derivative of Log$_{10}${$R(T)$} for the temperature-dependent resistance presented in Fig. 3 (d) is plotted in Fig. 3 (e-f) and fitted with Gaussians. The center and width of the Gaussian fits define $T_{IMT}$ ($T_{MIT}$) and $\Delta T_{Tr}$, respectively[42]. A distinct shift towards the right is observed in the Gaussian plot for the heating cycle of the bilayer film (Fig. 3 (e)), signifying a major increase in the $T_{IMT}$ (81.6 °C) compared to the single-layer reference film (68.2 °C). This further elucidates that the optimum bilayer VO$_2$/VO$_2$ film achieves a one-order increase in resistance switching while elevating $T_{IMT}$ by nearly 14 °C in a completely doping-free manner. A similar increase in the $T_{MIT}$ during the cooling cycles (Fig. 3 (f)) of both the films is also observed with the bilayer film and single-layer reference film demonstrating a $T_{MIT}$ of 58.2 °C and 46.8 °C, respectively. The difference between the $T_{IMT}$ (heating cycle) and $T_{MIT}$ (cooling cycle) further defines $\Delta T_{Hys}$.

All the bilayer VO$_{2-x}$/VO$_2$ films oxidized for different $T_{oxd}$ were further characterized using Raman Spectroscopy, as illustrated in Fig. 4 (a). The films oxidized for $T_{oxd}$ = 10 s and 12 s already show several strong VO$_2$ peaks, signifying the possibility of a nearly complete conversion from the elemental metallic V phase to the VO$_2$ phase. As discussed previously, this fast oxidation time may be ascribed to the oxidation of the top VO$_{2-x}$ film occurring from both the top, due to the atmosphere, and the bottom, due to the buffer layer. Weak V$_2$O$_5$ peaks start to appear at $T_{oxd}$ = 15 s and 24 s at 709.24 cm$^{-1}$ and 996.73 cm$^{-1}$. However, we still observe the VO$_2$ peaks dominating the entire oxidation range, both in terms of number and intensity compared to the competing V$_2$O$_5$ peaks. To contrast this, the Raman characterization for all the single-layer VO$_{2-x}$ reference films is presented in Fig. 4 (b). The pure V film ($T_{oxd}$ = 0 s) does not show any Raman peaks due to its pure metallic phase. The reference film oxidized for $T_{oxd}$ = 10 s only demonstrates four VO$_2$ peaks at 150.12 cm$^{-1}$, 201.52 cm$^{-1}$, 227.09 cm$^{-1}$, and 616.86 cm$^{-1}$. With an increase in oxidation times, more VO$_2$ peaks start appearing. The films with $T_{oxd}$ = 20 s, 25 s, and 30 s demonstrate only VO$_2$ peaks, except the sample with $T_{oxd}$ = 30 s, which begins to show signatures of weak V$_2$O$_5$ peaks. At $T_{oxd}$ = 40 s, a significant increase in V$_2$O$_5$ content as well as the emergence of a mixed phase of V$_2$O$_5$ and VO$_2$ can be observed. This monotonic progression is in agreement with previous reports, which attribute this behavior to the kinetics involved in the oxidation process[36].

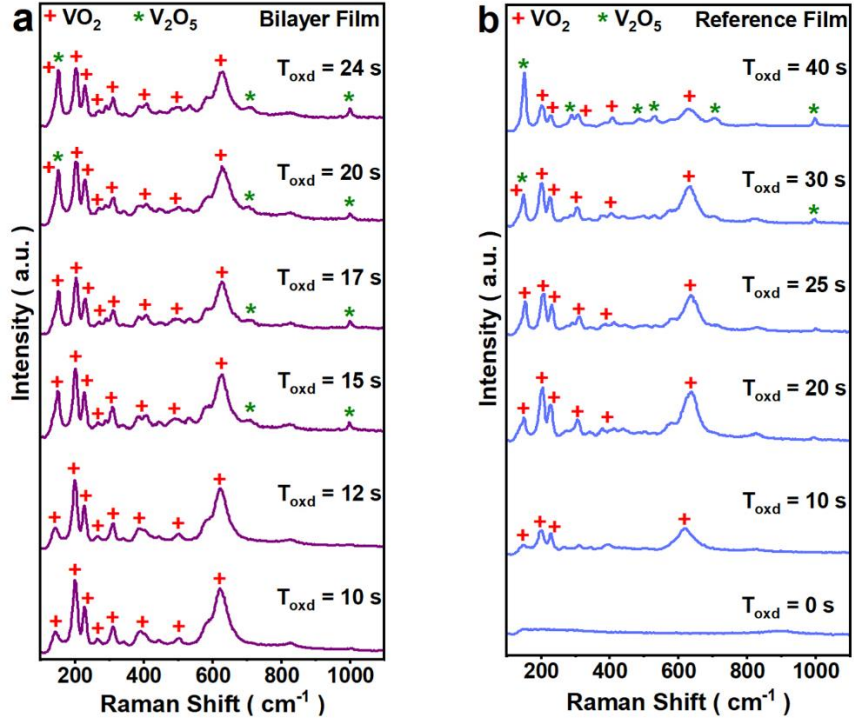

FIG. 4. Raman spectroscopy data recorded for all the films oxidized for different $T_{oxd}$ for both: (a) $VO_{2-x}/VO_2$ bilayer films; and (b) $VO_{2-x}$ single-layer reference films.

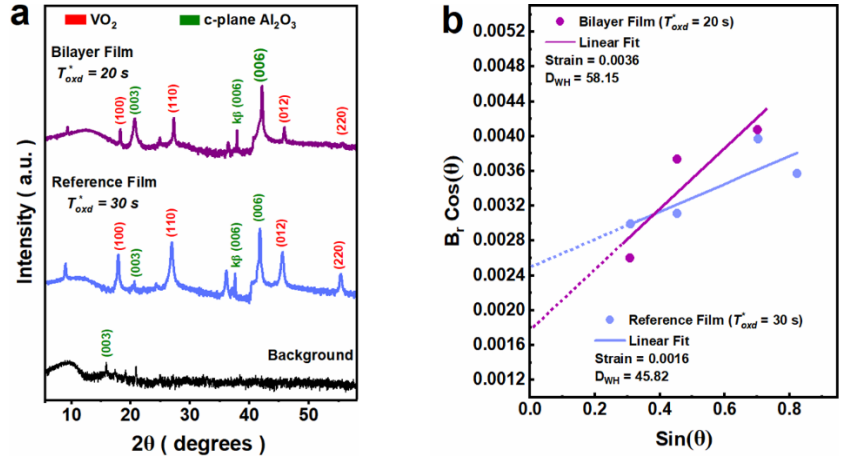

FIG. 5. (a) XRD data recorded for the optimum bilayer film ($T^*_{oxd} = 20$ s) and the optimum single-layer reference film ($T^*_{oxd} = 30$ s). (b) Linear fitting of $B_r \cdot \cos(\theta)$ vs $\sin(\theta)$ to extract the strain from the XRD characterization of the samples using the W-H formula (Eq. 2).

TABLE I. Average particle sizes obtained from XRD characterization of the samples.

| Sample | Peak Centre | Crystal Plane | FWHM | Scherrer Formula | | Williamson-Hall Formula | |
|---|---|---|---|---|---|---|---|
| | | | | Crystallite Size | Average Crystallite Size | Average Crystallite Size | Strain |
| Reference Film $T^*_{oxd} = 30\ s$ | 18.04 ° | (100) | 0.18 | 44.68 nm | 37.25 nm | 45.82 nm | 0.0016 |
| | 27.04 ° | (110) | 0.20 | 36.03 nm | | | |
| | 44.68 ° | (012) | 0.32 | 35.02 nm | | | |
| | 55.36 ° | (220) | 0.36 | 34.44 nm | | | |
| Bilayer Film $T^*_{oxd} = 20\ s$ | 17.92 ° | (100) | 0.18 | 44.68 nm | 41.09 nm | 58.15 nm | 0.0036 |
| | 26.94 ° | (110) | 0.24 | 40.85 nm | | | |
| | 44.56 ° | (012) | 0.26 | 40.51 nm | | | |
| | 55.29 ° | (220) | 0.31 | 40.34 nm | | | |

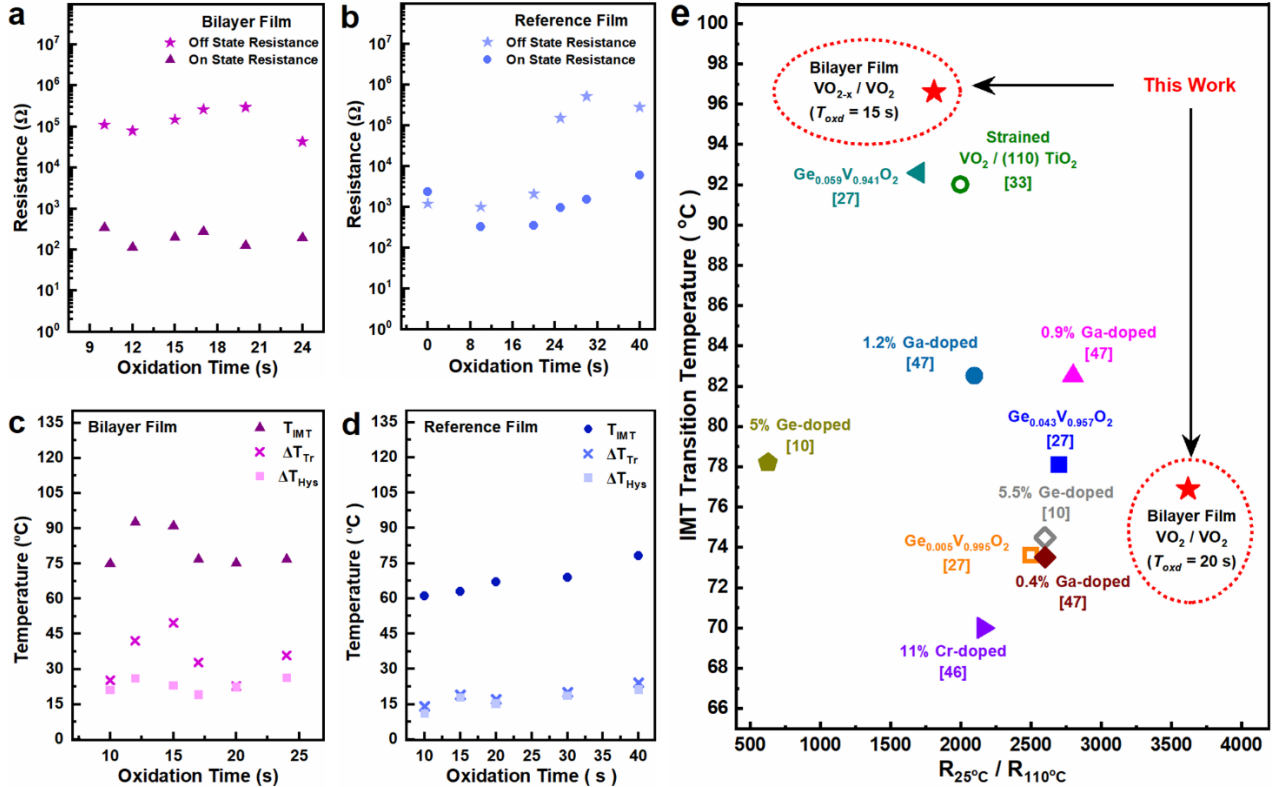

FIG. 6. Metallic on-state and insulating off-state resistance values as a function of oxidation times for: (a) Bilayer film, and (b) Single-layer reference film. IMT properties such as $T_{IMT}$, $\Delta T_{Tr}$, and $\Delta T_{Hys}$ as a function of oxidation times for: (c) Bilayer film, and (d) Single-layer reference film. (e) Comparison of the bilayer $VO_{2-x}/VO_2$ film and optimum $VO_2/VO_2$ film presented in this paper with previous reports on doped-$VO_2$ films with elevated transition temperatures.

The XRD plots for the optimum ($T^*_{oxd}$) bilayer as well as the single-layer reference film, are presented in Fig. 5 (a). $VO_2$ peaks are observed at 17.92°, 26.94°, 45.64°, and 55.36° in both the bilayer and single-layer films. These correspond to the (100), (110), (012), and (220) $VO_2$ planes, respectively. The resultant average particle sizes for both samples are presented in

Table I. The bilayer film and the single-layer reference film demonstrate crystallite sizes of 41.09 nm and 37.25 nm, respectively, using the Scherrer formula (Eq. 1) and crystallite sizes of 58.15 nm and 45.82 nm, respectively, using the W-H formula (Eq. 2). Slight deviations are often observed in the crystallite sizes extracted using these formulas. However, in both cases, a clear increase in grain size is observed in the bilayer film. Moreover, these sizes agree with previous reports for $VO_2$ films of a similar thickness[34]. To extract the strain in the films, the line fits for the $B_r \cdot \cos(\theta)$ vs $\sin(\theta)$ plots for both the bilayer and the single-layer films are presented in Fig. 5 (b). A significantly greater strain is observed in the bilayer film with $\eta = 0.0036$, while the single-layer reference film demonstrated an $\eta = 0.0016$.

Apart from the close to one-order increase in IMT resistance switching, the bilayer film also demonstrates significantly lower on-state resistance in its metallic phase as compared to the single-layer reference film. An on-state resistance of ~100 Ω was observed for the optimum bilayer film ($T^*_{oxd}$ = 20 s), which is one order lower than the on-state resistance observed for the single-layer reference film ($T^*_{oxd}$ = 30 s) (Fig. 6 (a-b)). This may be attributed to the crystallinity of the $VO_2$ buffer layer, which provides a better virtual substrate for the growth of the top $VO_2$ film as compared to c-plane sapphire due to improved lattice constant matching and higher crystallinity, resulting in the growth of a high-quality quasi-homogenous bilayer $VO_2/VO_2$ film with larger grain sizes. In addition, all the bilayer films oxidized at various other $T_{oxd}$ also show low on-state resistance values, as presented in Fig. 6 (a). This radically differs from the typical resistance trends observed in the single-layer reference films presented in Fig. 6 (b). At low $T_{oxd}$, due to the presence of unoxidized V in these single-layer reference films, they demonstrate low on-state resistance due to the metallic character of V. Following this, at $T^*_{oxd}$ and higher $T_{oxd}$, a gradual increase in both, the on-state and off-state resistance is observed which is attributed to the increase in oxide content. In contrast to this, the on-state and off-state resistance values of the bilayer film only show a slight variation with increasing $T_{oxd}$ values.

The plots illustrating the variation of $T_{IMT}$, $\Delta T_{Tr}$, and $\Delta T_{Hys}$ with different $T_{oxd}$ are illustrated in Fig. 6 (c-d). The highest $T_{IMT}$ achieved in bilayer $VO_{2-x}/VO_2$ films was 96.9 °C ($T_{oxd}$ = 15 s), which is one of the highest transition temperatures achieved in $VO_2$ films to date and is also demonstrated in a completely doping-free manner. The optimum bilayer $VO_2/VO_2$ film ($T^*_{oxd}$ = 20 s) also exhibited a high $T_{IMT}$ of 81.6 °C, which is notably higher than that of the optimum reference film (68.3 °C), while also demonstrating higher IMT resistance switching. This significant increase in the transition temperature may be attributed to the increased strain observed in the bilayer film relative to the single-layer reference film extracted via the W-H method (Eq. 2) in the previous section (Fig. 5 (a-b)). Based on prior reports, an elevation in transition temperature, when accompanied by increased strain in a $VO_2$ film, often suggests the existence of compressive strain along the c-axis of the film[38]. As experimentally demonstrated by Y. Muraoka et al.[38] with a $VO_2$ film grown on a $TiO_2$ (001) buffer layer, a compressive strain developed along the c-axis, which induced an increase in VO$_2$ transition temperatures. This strain was epitaxial in nature and was ascribed to the lattice parameter mismatch between the buffer layer and the VO$_2$ film[38, 43]. However, contrary to this, in the bilayer sample, the c-axis compressive strain is possibly non-epitaxial in nature due to its significant film thickness and may be attributed to the evolution of stoichiometry of the top VO$_{2-x}$ film during its APTO. This is further supported by the fact that the elevated transition temperatures are oxidation duration-dependent, indicating that the strain in the films is driven by $T_{oxd}$ and depends on the non-stoichiometric nature of the top VO$_{2-x}$ film. By adjusting the oxidation duration, the methodology presented in this paper enables the tuning of transition temperatures between roughly 74 °C and 97 °C, which is much higher than the original transition temperature of VO$_2$ at 68 °C. In addition, as presented in Fig. 6 (e), the bilayer samples displayed impressive performance when contrasted with prior reports regarding high transition temperature VO$_2$ films, which were predominantly accomplished by doping VO$_2$ with elements such as Cr[46], Ga[47], and Ge[27].

To further probe the effect of the buffer layer on the IMT resistance switching characteristics of the bilayer VO$_{2-x}$/VO$_2$ films, the resistance vs temperature plots for all the bilayer samples oxidized for different $T_{oxd}$ are presented in Fig. 7 (a-f). Under-oxidized bilayer films with $T_{oxd}$ = 10 s, 12 s, and 15 s demonstrate a significant change in the slope during their heating and cooling cycles. This is similar to the step-like feature reported by D. Lee et al.[43] during the structural phase transition in their VO$_{2-\delta}$/VO$_2$ bilayer system grown on TiO$_2$ using PLD. Here, VO$_{2-\delta}$ represented the oxygen-deficient top vanadium dioxide film with intrinsic point defects, fabricated with the goal of achieving a lower $T_{IMT}$ as compared to their ideal VO$_2$ bottom layer. Previously, C. H. Griffiths et al.[44] experimentally proved that oxygen-deficient VO$_{2-\delta}$ films have lower transition temperatures due to higher electron concentration as compared to perfectly stoichiometric VO$_2$ films. As a result of this difference between the $T_{IMT}$ of the top VO$_{2-\delta}$, and the bottom VO$_2$, D. Lee et al.[43] reported the existence of a transitional monoclinic/rutile heterostructure during the heating cycle, which stabilized a non-equilibrium metallic monoclinic phase responsible for the two-step structural phase transition.

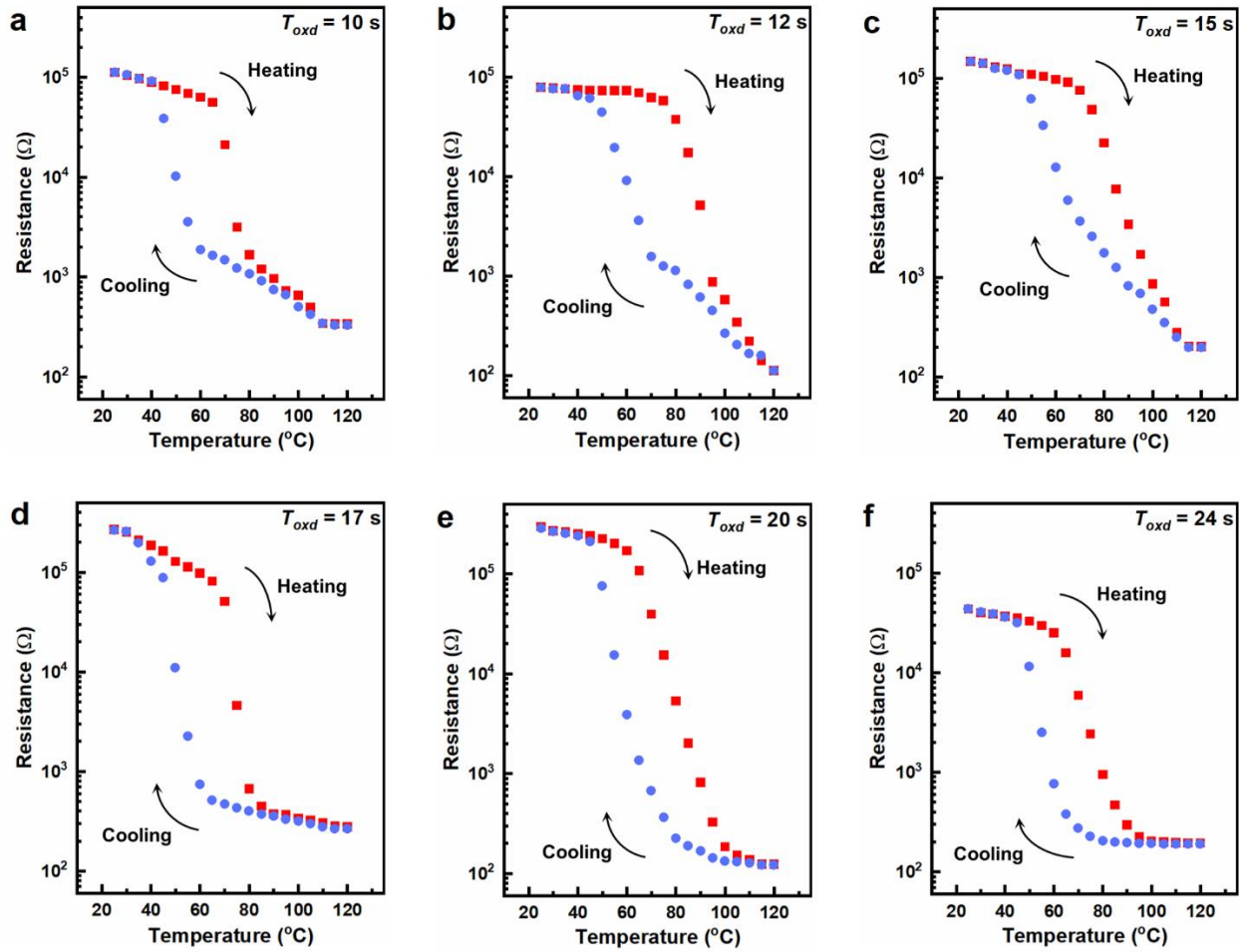

FIG. 7. Temperature-dependent resistance switching between 25 °C and 120 °C for all the bilayer films with varying oxidation times: (a) $T_{oxd}$ = 10 s, (b) $T_{oxd}$ = 12 s, (c) $T_{oxd}$ = 15 s, (d) $T_{oxd}$ = 17 s, (e) $T_{oxd}$ = 20 s, and (f) $T_{oxd}$ = 24 s.

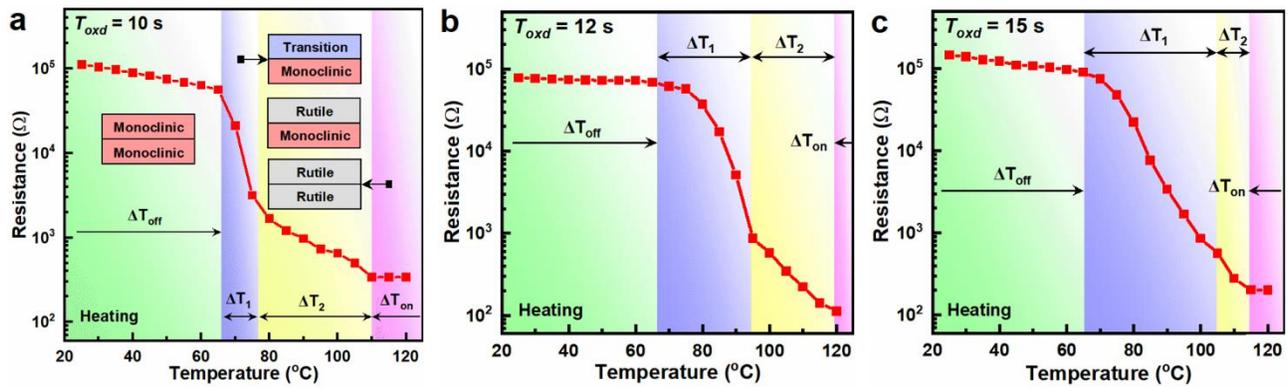

FIG. 8. Two-step IMT resistance transition illustrated in bilayer $VO_{2-x}/VO_2$ films with: (a) $T_{oxd}$ = 10 s (b) $T_{oxd}$ = 12 s, and (c) $T_{oxd}$ = 15 s. The inset graphically illustrates the formation of a transitional rutile/monoclinic quasi-heterostructure.

Reasoning along these lines, for the bilayer VO$_{2-x}$/VO$_2$ films wherein the top layer was oxidized for less than the optimum duration ($T_{oxd}$ = 10 s, 12 s, and 15 s), we can presume that the $T_{IMT}$ of the top VO$_{2-x}$ film will be lower than the $T_{IMT}$ of the buffer VO$_2$ film. Hence, the top film may undergo an earlier phase transition, resulting in the formation of a rutile/monoclinic heterostructure during the heating cycle, with the top VO$_{2-x}$ film existing in a transitional or rutile phase, while the buffer VO$_2$ film still exists in the insulating monoclinic phase as illustrated in Fig. 8 (a). This quasi-hetero state may give rise to a transitional parallel resistor system where the top VO$_{2-x}$ film and the buffer VO$_2$ film have substantially different resistivities at the same point of time during the heating cycle and may be considered as two resistors connected in parallel due to the configuration of the four-probe setup[39-41]. Hence, the change in slope observed in the temperature-dependent resistance of the bilayer films may be ascribed to this transitional quasi-hetero bilayer formed. This is also in agreement with the IMT resistance switching curves observed for the bilayer films with $T_{oxd}$ = 17 s, 20 s, and 24 s, where we see the change in slope disappear and observe a hysteresis pattern closer to that of a typical VO$_2$ film. As the $T_{oxd}$ gradually increases, complete oxidation of the top VO$_{2-x}$ film takes place, resulting in the top film and the buffer film having comparable $T_{IMT}$ and thus demonstrating a smooth typical IMT resistance transition. This is further graphically illustrated in Fig. 8 (a-c), where we see the widths of $\Delta T_1$ and $\Delta T_2$ monotonically increase and decrease respectively with increasing $T_{oxd}$, until the two merge and the change in slope disappears completely at $T_{oxd}$ = 17 s.

For future investigations aimed at simultaneously enhancing both IMT resistance switching and transition temperatures, several approaches may be considered to further modify the VO$_{2-x}$/VO$_2$ bilayer films introduced in this study. The fabrication of these bilayer films could be explored on different crystalline substrates, including TiO$_2$, SnO$_2$, and CeO$_2$, with different crystal orientations to introduce additional strain effects. Furthermore, the fabrication of bilayer films with sophisticated methods that have reported some of the highest orders of switching, such as MBE[14-16] and PLD[17-18], might yield an increased magnitude of switching as compared to single-layer films with identical thicknesses. Achieving higher $T_{IMT}$ is a considerable challenge as one will have to fabricate bilayer films with extremely high strain that is preserved. A possible future direction to surmount this challenge is to fabricate a tri-layer VO$_{2-x}$/VO$_{2-y}$/VO$_2$ film or a multi-layered VO$_2$ film, where *x* and *y* denote the non-stoichiometric nature of the films, by sequentially depositing each of the distinct oxide layers while maintaining a constant overall film thickness. By progressively increasing the number of buffer layers and maintaining a constant overall film thickness, the resulting structure begins to resemble a graded film commonly employed to preserve strain in thick films. This graded film structure may serve as an interesting future direction of study as increasing the number of buffer layers necessitates the optimization of several parameters, including the optimal number of buffer layers, the thickness of individual films, and the optimal oxidation conditions. Studying the exact layer-by-layer structural phase transition, their correlations, and their time scales in a multi-layered strongly-correlated oxide structure may be of substantial interest. However, each additional layer would severely increase the complexity of the system, both with respect to fabrication as well as the overall structural phase transition. Furthermore, the continuous reduction in the thicknesses of individual buffer layers would significantly enhance epitaxial effects, thus further contributing to its complexity.

With respect to the bilayer films reported in this paper, the investigation of the exact structural phase transition occurring at the $VO_{2-x}/VO_2$ quasi-heterogenous interface may also be of significant interest. In bulk $VO_2$ films, a triple-phase point was previously reported in the form of a third intermediate monoclinic phase (*M2*) in between the well-established insulating monoclinic phase (*M1*) and the metallic rutile phase (*R*)[45]. This intermediate *M2* phase, often called the correlated metallic monoclinic phase, was previously stabilized with the help of a bilayer system[43]. Several other reports investigating these intermediate phases in $VO_2$ have also relied on the study of bilayer films with the introduction of different buffer layers[34]. Thus, the $VO_{2-x}/VO_2$ bilayer films introduced in this paper may also act as a sandbox for investigating and stabilizing correlated intermediate crystal phases in vanadium dioxide and other Mott insulators.

## III. CONCLUSION

In this work, we introduced and studied the fabrication of a bilayer vanadium dioxide film with a self-buffer to elevate transition temperature while also increasing IMT resistance switching and decreasing on-state metallic resistance. Close to a one-order increase in IMT resistance switching was observed, with the optimum bilayer $VO_2/VO_2$ film demonstrating ~3.6 orders of resistance switching compared to the single-layer reference film. The introduction of the $VO_2$ buffer film also resulted in a one-order decrease in the on-state resistance. These features were attributed to the improved overall film quality and larger grain sizes due to the buffer $VO_2$ film acting as a highly suitable crystalline virtual substrate. A doping-free increase in transition temperature was also achieved with the under-oxidized bilayer $VO_{2-x}/VO_2$ film having $T_{oxd}$ = 15 s, demonstrating a $T_{IMT}$ of 96.9 °C, almost a 30 °C increase from the conventional transition temperature of $VO_2$. The oxidation time served as a knob to fabricate $VO_{2-x}/VO_2$ films with varying elevated transition temperatures, making APTO an ideal methodology for the fabrication of $VO_2$ thin film devices for applications requiring elevated $T_{IMT}$. An intriguing change in slope was also observed in the temperature-dependent resistance hysteresis curves of the bilayer $VO_{2-x}/VO_2$ samples, wherein the top $VO_{2-x}$ film was under-oxidized. Consequently, a quasi-heterostructure comprising a rutile/monoclinic interface was formed with the top film and the bottom film having different $T_{IMT}$. This heterogeneity manifests as a significant change in slope in the heating and cooling cycles of the temperature-dependent resistance plots. The results of this study should facilitate the adoption of bilayer $VO_2/VO_2$ films fabricated via APTO for applications requiring doping-free elevated transition temperatures, high magnitude of IMT resistance switching, and a limited thermal budget. Additionally, the two-step resistance transition demonstrated in the bilayer $VO_{2-x}/VO_2$ films may also form the basis of future works investigating the correlated intermediate phases in Mott insulators.

## ACKNOWLEDGMENTS

This project was supported by the Science and Engineering Research Board (SERB) India via Core Research Grant No. CRG/2022/005421 and used the Materials Science and Engineering (MSE) Raman characterization facility and the Advanced Center for Materials Science (ACMS) XRD facility.

## AUTHOR DECLARATIONS

The authors have no conflicts to disclose.